\documentclass[fleqn,usenatbib]{mnras}

\usepackage{newtxtext,newtxmath}
\usepackage[T1]{fontenc}

\DeclareRobustCommand{\VAN}[3]{#2}
\let\VANthebibliography\thebibliography
\def\thebibliography{\DeclareRobustCommand{\VAN}[3]{##3}\VANthebibliography}

\usepackage{graphicx}	% Including figure files
\usepackage{amsmath}	% Advanced maths commands
\title[Stochastic SFH in Dwarfs]{Modelling Stochastic Star Formation History of Dwarf Galaxies in GRUMPY}

\author[Pan \& Kravtsov]{Yue Pan,$^{1,2}$\thanks{E-mail: yue.pan@princeton.edu}
    Andrey Kravtsov$^{1,3,4}$\thanks{E-mail: kravtsov@uchicago.edu}
\\
\\
$^{1}$Department of Astronomy \& Astrophysics, The University of Chicago, Chicago, IL 60637, USA\\
$^{2}$Department of Astrophysical Sciences, Princeton University, Princeton, NJ 08544, USA\\
$^{3}$Enrico Fermi Institute, The University of Chicago, Chicago, IL 60637 USA\\
$^{4}$Kavli Institute for Cosmological Physics, The University of Chicago, Chicago, IL 60637 USA\\
}

\date{Accepted XXX. Received YYY; in original form ZZZ}

\pubyear{2022}
\begin{document}
\label{firstpage}
\pagerange{\pageref{firstpage}--\pageref{lastpage}}

%----------ABSTRACT
\maketitle
\begin{abstract}
We investigate the impact of bursty star formation on several galaxy scaling relations of dwarf galaxies using the \texttt{GRUMPY} galaxy formation model. While this model reproduces the star formation rate (SFR)--stellar mass, stellar mass--gas mass, and stellar mass--metallicity relations, the scatter of these relations in the original model is smaller than observed. We explore the effects of additional stochasticity of SFR on the scaling relations using a model that reproduces the level of SFR burstiness in high-resolution zoom-in simulations. The additional SFR stochasticity increases the scatter in the SFR--stellar mass relation to a level similar to that exhibited by most nearby dwarf galaxies. The most extreme observed starbursting dwarfs, however, require higher levels of SFR stochasticity.  We find that bursty star formation increases the scatter in the colour--magnitude distribution (CMD) for brighter dwarf galaxies $(M_V < -12)$ to the observed level, but not for fainter ones for which scatter remains significantly smaller than observed. This is due to the predominant old stellar populations in these faint model galaxies and their generally declining SFR over the past 10 Gyrs, rather than quenching caused by reionization. We examine the possibility that the colour scatter is due to scatter in metallicity, but show that the level of scatter required leads to an overestimation of scatter in the metallicity--mass relation. This illustrates that the scatter of observed scaling relations in the dwarf galaxy regime represents a powerful constraint on the properties of their star formation. 

\end{abstract}

%----------Keywords

%%%%%%%%%%%%%%%%% BODY OF PAPER %%%%%%%%%%%%%%%%%%

%----------SECTION 1: Introduction
\section{Introduction}
Scaling relations between galaxy properties are a valuable window into processes driving galaxy formation \citep{Kennicutt.1989, Kauffmann.etal.2003, Kormendy.etal.2009, Lilly.etal.2013, Tacchella.etal.2016,Mason.etal.2023}.  Over the past two decades, the relation between galaxy stellar mass, $M_\star$, and its star formation rate, $\rm SFR$, was studied very actively \citep[see, e.g.,][and references therein]{Popesso.etal.2023}. The nearly linear relation, often referred to as the star-forming main sequence (SFMS), is widely considered to be instrumental to understanding how stellar mass of galaxies is built up and the processes that regulate star formation in galaxies \citep[e.g.,][]{Leitner.Kravtsov.2011,Leitner.2012,Lee.etal.2015}. 

Observations revealed that the SFMS exists both at $z=0$ \citep[e.g.,][]{ Brinchmann.etal.2004, Salim.etal.2007, Renzini.Peng.2015, Cano-Diaz.etal.2016} and at higher redshifts \citep[see e.g., ][]{Noeske.etal.2007,Daddi.etal.2007, Pannella.etal.2009, Rodighiero.etal.2010, Wuyts.etal.2011, Whitaker.etal.2012,  Mancuso.etal.2016, DSilva.etal.2023}.

In general, both the form of the SFMS (its amplitude and slope) and the scatter of the
$\log_{10}\rm SFR$ around the mean relation, $\Delta_{\rm{MS}}$, are important. The scatter is directly related to seemingly random variations (stochasticity) of star formation rate and can provide valuable insights into the physical processes that affect star formation histories (SFHs) on different mass scales \citep[e.g.,][]{Matthee.Schaye.2019}. 

For bright galaxies,  $\Delta_{\rm{MS}}\approx 0.20-0.35$ dex is estimated in both observations and simulations \citep[e.g.,][]{Daddi.etal.2007, Whitaker.etal.2012,Speagle.etal.2014,Torrey.etal.2014, Sparre.etal.2015,Matthee.Schaye.2019}. Moreover, it has been found that the scatter $\Delta_{\rm{MS}}$ of the SFMS can be accurately described by the log-normal distribution \citep[e.g.,][]{Guo.etal.2013, Chang.etal.2015, Schreiber.etal.2015, Davies.etal.2019}. 
There are also indications that scatter increases with decreasing stellar mass of galaxies \citep[e.g.,][]{Santini.etal.2017} and that it is particularly prevalent in galaxies at $z\gtrsim 6$ \citep{Dressler.etal.2023,Looser.etal.2023,Atek.etal.2023,Endsley.etal.2023,Ciesla.etal.2023}. 

Some studies have used the fact that $\Delta_{\rm MS}$ is moderate to suggest that galaxies' SFRs on the MS persist over long periods, regulated by various physical processes such as gas accretion and feedback-driven outflows \citep{Bouche.etal.2010, Daddi.etal.2010, Genzel.etal.2010, Dave.Oppenheimer.Finlator.2011, Lilly.etal.2013, Dekel.Mandelker.2014,Rodriguez-Puebla.etal.2016, Tacchella.etal.2016}. However, there is no consensus on the redshift evolution of $\Delta_{\rm{MS}}$. While some studies concluded that it is independent of redshift \citep[e.g.,][]{Daddi.etal.2007, Noeske.etal.2007, Whitaker.etal.2012, Speagle.etal.2014, Ciesla.etal.2014, Pessa.etal.2021}, others find that it evolves with redshift \citep[e.g.,][]{Kurczynski.etal.2016,Santini.etal.2017,Katsianis.etal.2019,Tacchella.Forbes.Caplar.2020,Davies.etal.2022,Shin.etal.2023}. 

Building on these observations and models, \citet{Caplar.Tacchella.2019} utilized 
the power spectral density (PSD) distribution to characterize the stochasticity of SFHs for all star-forming galaxies with $M_\star \approx 10^{10} M_\odot$. They found that the SFHs of galaxies tend to be correlated on a timescale of $\approx 200$ million years. This timescale is shorter than the dynamical time of the dark matter halo, implying that the baryonic effects, such as feedback and reincorporation of galactic winds, which operate on the dynamical timescale of galaxies, play a crucial role in shaping the SFHs of galaxies. 
In several follow-up studies, the PSD was used to investigate the effect of gas cycling through ISM and giant molecular clouds (GMCs) on the SFR of galaxies \citep{Tacchella.Forbes.Caplar.2020}, analysis of the level of SFR stochasticity in different galaxy formation models and simulations \citep{Iyer.etal.2020}, and the impact of different observational SFR tracers on the inferred SFR stochasticity \citep{Iyer.etal.2022}. These studies elucidated the contribution of different physical processes to the variation of SFR on different time scales. 

On short time scales ($\lesssim 100$ Myr), the amplitude of the SFR fluctuations contains valuable information about the creation and destruction of individual GMCs \citep{Scalo.Struck-Marcell.1984, Scalo.Struck-Marcell.1986, Kruijssen.etal.2014, Krumholz.Kruijssen.2015,Semenov.etal.2017, Faucher.Giguere.2018, Orr.Hayward.Hopkins.2019} caused by supernova explosions, cosmic rays, and photoionization \citep{Gnedin.Kravtsov.Chen.2008, Parrish.etal.2009, Hopkins.etal.2014, Faucher.Giguere.2018, Tacchella.Forbes.Caplar.2020,Semenov.etal.2021a,Semenov.etal.2021b}. In the regime of extremely low SFR in dwarf galaxies, the stochasticity can even be caused by the formation of individual massive stars \citep[see e.g.][]{Fumagalli.etal.2011, da.Silva.etal.2012,daSilva.etal.2014}. 

On intermediate time scales of $\sim 0.1-1$ Gyr, various dynamic processes such as galactic winds, disc formation, bulge growth, environmental effects, bar-induced inflows, and galaxy mergers are thought to influence the fluctuations in star formation \citep{Oppenheimer.Dave.2008, Tacchella.etal.2016, Sparre.etal.2017, Torrey.etal.2018, Wang.Lilly.2020a,Wang.Lilly.2020b}. 

On longer time scales ($> 1$Gyr), galaxy quenching and the overall interplay between gas accretion and outflows are known to play significant roles \citep{Behroozi.etal.2013, Moster.etal.2018, Birrer.etal.2014, Rodriguez-Puebla.etal.2016,Weinberger.etal.2017, Angles.Alcazar.etal.2017, Tacchella.etal.2018, Behroozi.etal.2019}.

Although the stochasticity of SFR in massive galaxies was explored extensively, our understanding of the scatter in the SFMS in the dwarf galaxy regime remains rather limited, even though studies of star formation in nearby dwarf galaxies indicate that their SFRs are bursty \citep[e.g.,][]{Emami.etal.2019}. While the approximately linear relation between SFR and stellar mass extends to galaxies of stellar mass as low as $\approx 10^5 M_\odot$, the scatter growth dramatically with decreasing stellar mass for galaxies with $M_\star\lesssim 10^8\, M_\odot$ \citep[see, e.g.,][]{Kravtsov.Manwadkar.2022}. In particular, the latter authors showed that scatter stemming from the variations of mass assembly histories and structural properties of dwarf galaxies is not sufficient to explain the observed scatter of specific SFRs in the dwarf galaxy regime. 

This is not entirely surprising, given that star formation in dwarf galaxies is modulated by a few individual star forming regions, which leads to burstier star formation. At the same time, outflows are expected to be more efficient in dwarf galaxies which may result in periods of low SFR following bursts thereby leading to stronger variations of SFR. These processes are confined within localized regions in galaxies and require detailed modeling of interstellar medium and star-forming regions in galaxies, as well as their destruction and accompanying feedback effects. 

However, many galaxy formation models and simulations do not model or resolve the relevant scales and model relevant processes. For example, \citet{Kravtsov.Manwadkar.2022} show that a simple regulator-based \textsc{}{GRUMPY} model reproduces the mean observed correlations between properties of dwarf galaxies, as well as luminosity function and radial distribution of Milky Way's dwarf satellites \citep{Manwadkar.Kravtsov.2022}. However, the model does not model small-scale star-forming regions and thus predicts a much smaller scatter than observed in the colour--magnitude and stellar mass--SFR relations.  Similarly, \citet{Pan.etal.2023} found that dwarf galaxy colours in the Auriga cosmological hydrodynamical simulations \citep{Grand.etal.2017,Grand.etal.2021} have a significantly smaller scatter than observed. This is likely due to the fact that ISM in these simulations is pressurized and ISM and formation and destruction of individual star-forming processes are not modeled, but overall star formation is modeled statistically with a subgrid model. 

The goal of this study is to evaluate the degree of SFR stochasticity in dwarf galaxies required to reproduce observed scatter in the colour-magnitude and stellar mass--SFR relations. To this end, we use the regulator-based \textsc{GRUMPY} galaxy formation model mentioned above, but add scatter to the SFR predicted by the model in a controlled manner. 

This paper is organized as follows. We describe  the \texttt{GRUMPY} model in Section~\ref{sec:model}. We summarize the PSD formalism used in the model of SFR stochasticity in Section~\ref{sec:PSD} and the revised model with stochasticity in Section~\ref{sec:grumpy_stoch_SFR}. We investigate the effects of stochasticity on the SFR--mass relation in Section~\ref{sec:effect_SFR}, the colour--magnitude diagram in Section~\ref{sec:effect_color}, and the metallicity--stellar mass relation in Section~\ref{sec:effect_feh}. In Section~\ref{sec:discussion}, we compare our results with results from other theoretical models and simulations in Section~\ref{sec:dis_comp_models}and~\ref{sec:dis_color_scatter}, explore the sensitivity of different SFR indicators to the stochasticity of SFR in observations in Section~\ref{sec:dis_sfr_indicators}, and discuss several caveats in Section~\ref{sec:caveats}. Finally, we present our conclusions in Section~\ref{sec:conclusion}.

The cosmological parameters adopted in this study are those of the Caterpillar simulation suite: $h=H_0/100=0.6711$, $\Omega_{\rm m0}=0.32$, $\Omega_\Lambda=0.68$.

%----------SECTION 2: Model stochastic SFR
\section{Galaxy formation model with stochastic SFR}
\label{sec:model}
%------------------------------------------

We begin with description of how stochasticity is modeled in this study in conjunction with the SFR produced by the fiducial \texttt{GRUMPY} model \citep{Kravtsov.Manwadkar.2022}. We first review the \texttt{GRUMPY} model with a focus on the modeling of star formation, molecular hydrogen, and metallicity -- quantities most relevant for the present study. We then describe the stochastic SFR framework based on the power spectral density approach of \citet{Tacchella.etal.2018}. Finally, we combine the two components and describe how stochastic SFR is modelled in this study.

%----------SUBSECTION 2.1: GRUMPY
\subsection{The \textsc{GRUMPY} galaxy formation model}
\label{sec:grumpy_description}
%----------SUBSUBSECTION 2.1.1: General formalism
\subsubsection{Halo mass tracks and GRUMPY galaxy formation model}
\label{sec:general_grumpy}
We model the population of dwarf galaxies around the Milky Way using tracks of haloes from the Caterpillar suite of $N$-body simulations \citep{Griffen.etal.2016}, which simulated 32 MW-sized haloes.\footnote{The Caterpillar project website can be found at \url{https://www.caterpillarproject.org}.} We use the highest resolution set of simulations, LX14, which allows to maximize the size of our model galaxy sample. 

The halo tracks were produced by a modified version of the Rockstar halo finder and the Consistent Trees Code \citep{Behroozi.etal.2013}, as described in Section 2.5 of \citet{Griffen.etal.2016}. As shown by \citet[][see their Fig. 1]{Manwadkar.Kravtsov.2022}, the subhalo peak mass function in the LX14 simulations is complete at $\mu=M_{\rm peak}/M_{\rm host} \gtrsim 4 \times 10^{-6}$ or $M_{\rm peak} \approx 4 \times 10^{6} M_\odot$ for the host halo mass $M_{\rm host}\approx 10^{12}\, M_\odot$, even in the innermost regions of the host ($r < 50$ kpc). This degree of completeness is sufficient to model the full range of luminosities observed in Milky Way satellites since the faintest ultra-faint dwarfs (UFDs) are hosted in haloes of $M_{\rm peak}\gtrsim 10^7\, M_\odot$ in our model \citep[][]{Kravtsov.Manwadkar.2022,Manwadkar.Kravtsov.2022}.

The halo mass evolution tracks from the Caterpillar suite are used as the basis for the \texttt{GRUMPY} galaxy formation model {Kravtsov.Manwadkar.2022}  based on a regulator-type galaxy formation framework \citep[e.g.,][]{Krumholz.Dekel.2012,Lilly.etal.2013,Feldmann.2013}. Namely, the model solves a system of coupled differential equations to follow the evolution of key galaxy properties, including the effects of UV heating after reionization, gas accretion suppression onto small mass haloes, galactic outflows, a model for the gaseous disk and its size, molecular hydrogen mass, star formation, and more. The model also accounts for the evolution of the half-mass radius of the stellar distribution. The galaxy model parameters used in this study are identical to those in the fiducial model of \citet{Manwadkar.Kravtsov.2022}.

As the primary of this study is on the stochasticity of star formation rate, we briefly review how star formation is modeled in the \texttt{GRUMPY} model. The model uses the evolved exponential radial gas profile, $\Sigma_g(R)$, to estimate the mass of molecular hydrogen using the model of  \citet{Gnedin.Draine.2014}: 
\begin{equation}
    M_{\rm H_2} = 2\pi\int_0^{\rm{R_{HI}}}f_{\rm{H_2}}(\Sigma_g)\Sigma_g(R)RdR
\end{equation}
where $R_{\rm{HI}}$ is the radius corresponding to the assumed self-shielding threshold and $f_{\rm H_2}$ is molecular fraction computed using current gas metallicity and mean star formation rate. 

The star formation rate is computed using $M_{\rm{H_2}}$ assuming a constant depletion time of molecular gas, $\tau_{\rm{dep, H_2}}$, and instantaneous recycling of the gas:
\begin{equation}
    \dot {\mathcal{M}}_\star = (1-\mathcal{R})\dot {M}_\star
\end{equation}
where 
\begin{equation}
    \dot {M}_\star = \frac{M_{\rm{H_2}}}{\tau_{\rm{dep, H_2}}} = \rm{SFR}
\end{equation}
$\mathcal{R}=0.44$ represents the proportion of gas that is converted into stars and subsequently returned to the ISM assuming the \citet{Chabrier.2003} initial mass function of stars \citep[e.g.,][]{Leitner.Kravtsov.2011, Vincenzo.etal.2016}. Meanwhile, $\tau_{\rm{dep, H_2}}$ is assumed to be $\tau_{\rm{dep, H_2}}=2.5$ Gyr in the fiducial model consistent with measurements in nearby galaxies \citep[e.g.,][]{Bigiel.etal.2008,Bigiel.Blitz.2012,Bolatto.etal.2011,Rahman.etal.2012,Leroy.etal.2013}. 

Observations indicate that $\tau_{\rm H_2}$ may vary between galaxies and with redshift. Thus, the assumption of constant $\tau_{\rm H_2}$ is one of the reasons the fiducial model may underestimate SFR stochasticity. 

\citet[][]{Kravtsov.Manwadkar.2022} and \citet{Manwadkar.Kravtsov.2022} showed that the model successfully reproduces the general trends observed in the SFR--mass relation, mass--metallicity relation, and colour--magnitude relation of Milky Way satellite galaxies, as well as luminosity function and radial distribution of the Milky Way satellites. However, as noted in the Introduction, the model predicts a significantly smaller scatter in star formation rates and colours of dwarf galaxies than what is actually observed in the Universe. 

This is not surprising because the \texttt{GRUMPY} framework does not include modeling of the processes that can result in significant stochasticity of star formation, such as formation and destruction of individual star-forming regions \citep[e.g.,][]{Iyer.etal.2020}. Here we introduce additional SFR stochasticity in a controlled manner using PSD formalism, as described below. However, we first describe the sources of stochasticity that {\it are} included in the model.

%----------SUBSECTION 2.3 Source of stochasticity in the model
\subsubsection{Sources of stochasticity in the original model}\label{sec:source_stoch}

The fiducial \texttt{GRUMPY} model includes two sources of stochasticity. 
The first arises from the variations of the mass assembly histories (MAHs) of haloes of a given final mass. The diversity in halo MAHs can result in a range of scatter in galaxy properties and can be particularly significant in the ultra-faint dwarf (UFD) regime due to the suppression of gas accretion and star formation after the epoch of reionization \citep{Rey.etal.2019, Katz.etal.2020}.

The second source of stochasticity arises from the assumed stochasticity in the size of gas distribution in galaxies. The latter is modeled using an exponential profile:
\begin{equation}
\Sigma_g(R) = \Sigma_0 \exp\left(-\frac{R}{R_d}\right)
\end{equation}
where $\Sigma_0$ is the central surface density, $R_d$ is the disc size. The model assumes that at each redshift $z$ during evolution, the disc size is proportional to the halo radius enclosing density contrast of 200 times the critical density, $R_{\rm{200c}}(z)$:
\begin{equation}
R_d = \chi_d R_{\rm{200c}}(z)
\end{equation}
The value of $\chi_d$ has a mean of 0.06 \citep{Mo.etal.1998} and a random log-normal Gaussian scatter of 0.25 dex to account for the fact that there is significant scatter in disc sizes at a given stellar mass in observations. 
The same random $\chi_d$ is used throughout its evolution for a given galaxy, but different galaxies have different $\chi_d$ values.

Although these sources of stochasticity between galaxies produce some scatter in galaxy scaling relations and in particular reproduce scatter in observed size--magnitude relation \citep{Kravtsov.Manwadkar.2022,Manwadkar.Kravtsov.2022}, the scatter in the $M_\star-\rm SFR$ and colour--magnitude relations is considerably smaller than in the corresponding observed relations, hence motivating additional sources of SFR scatter.

%----------SUBSECTION 2.2: PSD
\subsection{Power Spectral Density of stochastic SFR}
\label{sec:PSD}
We follow \citet{Caplar.Tacchella.2019} and model the specific star formation rate (SFR) of a galaxy with stellar mass $M_\star$ relative to the main sequence (MS) relation as:
\begin{equation}
\Delta_{\rm{MS}} = \log_{10}\left(\frac{\rm{SFR}}{\langle\rm{{SFR}}\rangle(M_\star),}\right)
\end{equation}
where $\langle\rm SFR\rangle(M_\star)$ represents the SFR of the mean MS at the mass $M_\star$ and $\Delta_{\rm{MS}}$ is a Gaussian random number. Consequently, the distribution of $\rm{SFR}/\langle\rm SFR\rangle(M_\star)$ is log-normal, which is consistent with observations \citep{Guo.etal.2013, Davies.etal.2019, Caplar.Tacchella.2019}.

\begin{figure*}
	\includegraphics[width=0.95\textwidth]{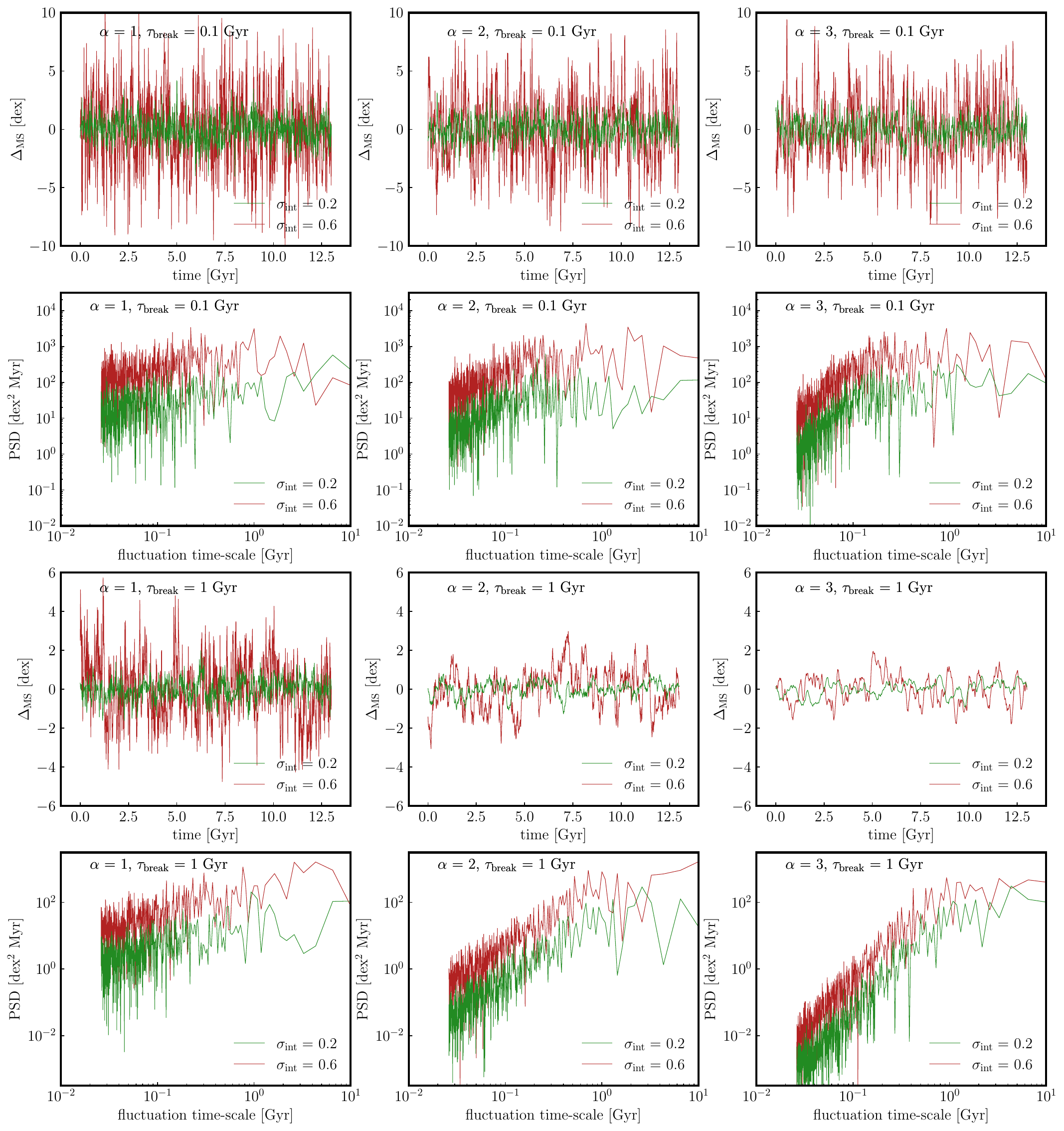}  
	\caption{An illustration of how $\Delta_{\rm{MS}}$ and its PSD depend on $\alpha$, $\tau_{\rm{break}}$, and $\sigma_{\rm{int}}$. We show $\alpha = 1, 2, 3$ and $\tau_{\rm{break}} = 0.1, 1$ Gyr. In each panel we show $\sigma_{\rm{int}} = 0.2$ (grey) and 0.6 (purple). As $\alpha$ increases and $\tau_{\rm{break}}$ increases, adjacent points become more correlated. A larger $\sigma_{\rm{int}}$ value elevates the PSD on all timescales.
} \label{fig:illustration_params}
\end{figure*}

%----------FIG: illustration of PSD parameters

We thus model stochasticity of star formation relative to the mean $M_\star-{\rm SFR}$ relation by producing  correlated Gaussian random number for each time step during galaxy evolution integration. The numbers are generated assuming a given power spectral density (PSD) of the time series in the Fourier domain.

The time series, represented by a discrete set of values $s(t_n) = s_n$ at evenly spaced intervals $t_n = t_{\rm min} + n\Delta t$ for $n=[0,N-1]$, covers a time span of $T = N\Delta t$. The Fourier transform of the series, $S_k$, where $k$ represents the wave number and ranges from $-N/2$ to $N/2$:
\begin{equation}
\begin{aligned}
    s_n & = \frac{1}{N}\sum\limits_{k=0}^{N-1} S_k\cdot \exp\left(\frac{i2\pi n}{N}k\right)\\
    & =\frac{1}{N}\sum\limits_{k=0}^{N-1} S_k\cdot \exp\left[\frac{i 2\pi (t_n-t_{\rm min})}{\Delta t N}k\right]\\
    & =\frac{1}{N}\sum\limits_{k=0}^{N-1} S_k\cdot \exp\left[i 2\pi (t_n-t_{\rm min})f_k\right]
\end{aligned}
\end{equation}
The DC signal has a frequency $f_0 = 0$ and a corresponding $S_0$ value of 1. The frequency $f_k$ is defined as $k/T$.

According to the Parseval's theorem, for a real series $s_n$:
\begin{equation}
\frac{1}{N}\sum\limits_{n=0}^{N-1} s^2_n = \frac{1}{N^2}\sum\limits_{k=0}^{N-1} \vert S_k\vert^2
\end{equation}
For a series $s_n$ with a mean of zero, the left-hand side represents the variance, and the right-hand side indicates that the sum of the squares of Fourier components of $s_n$ divided by $N^2$ should equal the variance of $s_n$.

The power spectrum $P(k)$ is defined as:
\begin{equation}
P(k) = \frac{1}{N^2} \vert S_k\vert^2
\end{equation}
Note that $P(k)$ is dimensionless, and according to the Parseval's theorem above:

\begin{equation}
\sigma^2(s_n) = \sum\limits_{k=0}^{N-1}P(k)
\end{equation}

The power spectral density (PSD) is defined as
\begin{equation}
{\rm PSD}(k) = \frac{(\Delta t)^2}{T} \vert S_k\vert^2 = \frac{T}{N^2} N^2 P(k) = T P(k) = N\Delta t P(k).
\end{equation}
and has unit of time.

To generate a sequence of Gaussian random numbers with a given power spectrum, we use a grid of Fourier components at frequencies $f_{\rm grid}$ with amplitudes set to random numbers drawn from the Gaussian pdf with zero mean and unit variance and multiplied by  $\sqrt{P(k)}=\sqrt{{\rm PSD}(f)/T}$. 

We follow  \citep{Caplar.Tacchella.2019} and use PSD of the form
\begin{equation}
    \rm{PSD}(\emph{f}) = \frac{\sigma^2}{1+(\tau_{\rm{break}}\ \emph{f}_{\rm{grid}})^{-\alpha}},
\end{equation}\label{eqn:PSD}
where $\sigma = \sigma_{\rm{int}}\tau_{\rm{decor}}$ characterizes the variability over a long time scale and  $\tau_{\rm{break}}=2\pi\tau_{\rm{decor}}$ characterizes the timescale over which the random numbers are no longer correlated.  

We generate a time series of $\Delta_{\rm MS}$ perturbations this way with $\Delta t=10\,\rm Myr$ -- the value chosen to correspond to the typical lifetime of star forming regions in galaxies \citep{Fall.etal.2010, Kruijssen.etal.2019, Semenov.etal.2017, Tacchella.Forbes.Caplar.2020, Semenov.etal.2021a}.

Figure~\ref{fig:illustration_params} shows an illustration of how $\Delta_{\rm{MS}}$ and its PSD changes with $\alpha$, $\tau_{\rm{break}}$, and $\sigma_{\rm{int}}$. In general, a larger $\alpha$ and $\tau_{\rm{break}}$ increases the correlation between adjacent points, and a larger $\sigma_{\rm{int}}$ increases the power on all timescales.

%----------FIG: stochastic SFR PSD example
\begin{figure*}
	\includegraphics[width=\textwidth]{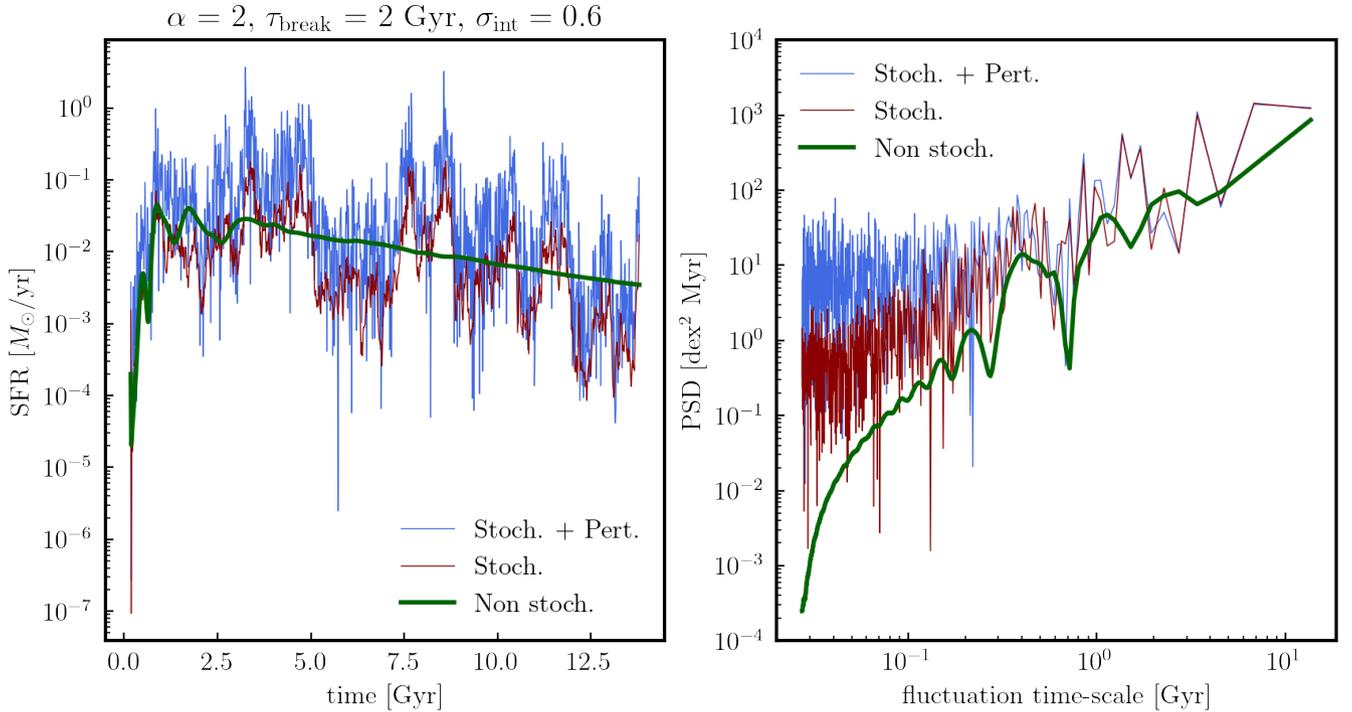}  
	\caption{An illustration of non-stochastic SFH, stochastic SFH, and stochastic+perturbed SFH (left) along with their corresponding PSD (right). The solid green line represents the SFH of a typical halo track that exhibits high star formation during the first 2 Gyr, followed by a steady SFR decline. Due to the rather smooth SFH, the power on short-term fluctuations (thick green line on the right) is low. To create a more bursty SFH, we add stochasticity to the SFH of this sample galaxy, following the process described in Section~\ref{sec:model}, which yields a red curve on the left. Since we manually introduce short-term fluctuations, the power on short-term fluctuations is much greater than in the non-stochastic case (red curve on the right compared to green curve). Moreover, as shown in Section~\ref{sec:effect_SFR}, we test another model where we introduce an additional PSD corresponding to a delta function on the timescale of $\sim 13$ Myr that increases SFR by a random number. This is a test to show how much additional perturbation is needed to match the high SFR of starburst galaxies like those in \citet{Lin.etal.2022}. As demonstrated, adding such a PSD results in significantly greater power on short fluctuation timescales (blue curve on the right).} \label{fig:PSD}
\end{figure*}

%----------FIG: SFR vs mass
\begin{figure*}
	\includegraphics[width=\textwidth]{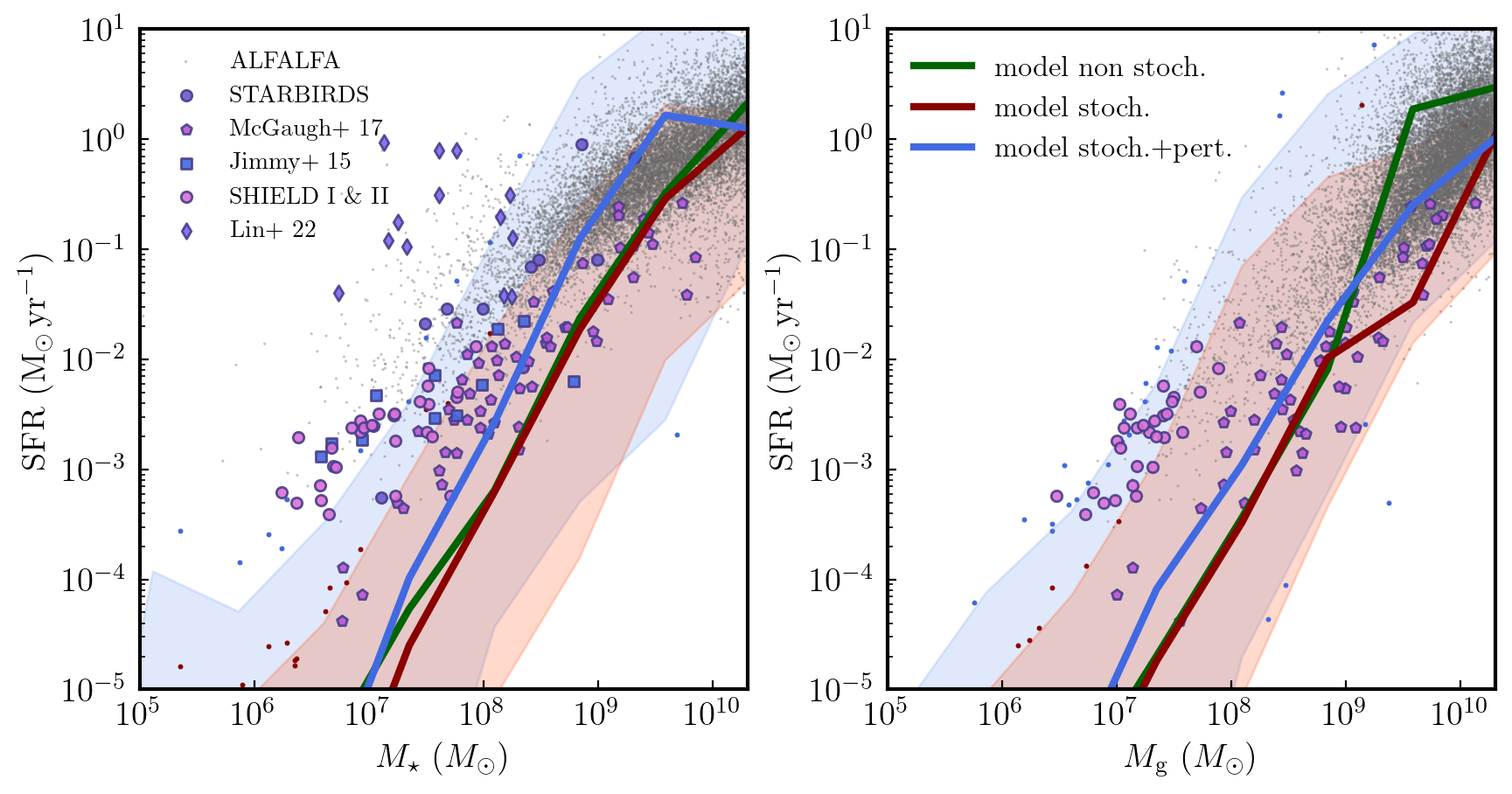}  
	\caption{$M_\star-\rm SFR$ (left panel) and $M_{\rm g}-\rm SFR$ (right panel) relations for model galaxies in the non-stochastic run (green) and stochastic run (red) as a comparison to observations such as the ALFALFA survey \citep[][grey dots]{Durbala.etal.2020}, STARBIRDS survey \citep[][purple circles]{McQuinn.etal.2019}, \citet[][pentagons]{McGaugh.etal.2017}, \citet[][squares]{Jimmy.etal.2015}, the SHIELD I \& II surveys \citep[][McQuinn et al. 2022, submitted; pink circles]{McQuinn.etal.2015a}, and \citet[][diamonds]{Lin.etal.2022}. For \citet{Lin.etal.2022}, we only include points that have SFR fractional error $<20$\%. The median SFR values in each mass bin for the non-stochastic run (green), stochastic run (red), and stochastic+post-process perturbation (blue) are shown by solid lines. The transparent colour band illustrates the 2-sigma variation, and any points outside the 2-sigma width of the median are plotted as individual points. We find that the linear relation between SFR and $M_\star$ extends down to $M_\star \sim 10^5 M_\odot$ dwarf galaxies. The 2-sigma width of the SFR-$M_\star$ and SFR-$M_g$ relation in the non-stochastic run is relatively smaller than that for the stochastic run. When we add additional perturbations to the model SFR using another form of PSD as described in Section~\ref{sec:effect_SFR}, we obtain a much higher SFR (blue) that is within the same range as observed galaxies in SHIELD I \& II, although we are unable to reach the same high SFRs as those in \citet{Lin.etal.2022}. As discussed in Section~\ref{sec:discussion}, different SFR indicators can cause significant differences in the SFR value, with indicators such as H$\alpha$ used in \citet{Lin.etal.2022} tending to produce higher values. Overall, our model results agree well with observations, with future studies warranted to explore the physics that cause star bursts.} \label{fig:SFR_mass}
\end{figure*}

%----------FIG: gas mass stellar mass
\begin{figure}
	\includegraphics[width=\columnwidth]{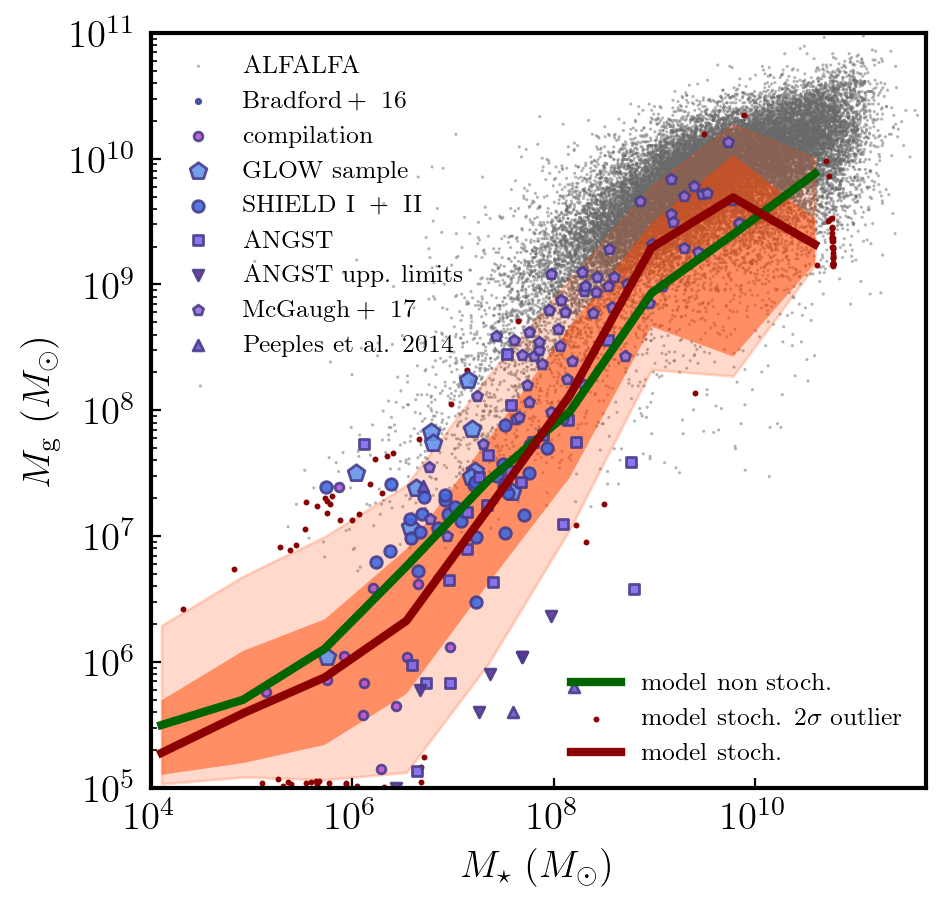}  
	\caption{$M_g$-$M_\star$ relation for model galaxies in the non-stochastic and stochastic runs compared with observations. Similar to Figure~\ref{fig:SFR_mass}, we plot the median SFR for each mass bin and the 1-sigma and 2-sigma width for the stochastic run. Switching from a non-stochastic to a stochastic SFR model does not significantly alter the median relationship between gas and stellar mass.
} \label{fig:gas_stellar_mass}
\end{figure}

%----------SUBSECTION 2.4: Model stochastic SFR in GRUMPY
\subsection{Modelling stochastic SFR in GRUMPY}\label{sec:grumpy_stoch_SFR}

In what follows, we use the PSD given by eq.~\ref{eqn:PSD} with the values $\alpha = 2$, $\tau_{\rm{break}} = 2$ Gyr, and $\sigma_{\rm{int}} = 0.6$. We find that using the PSD given by equation~\ref{eqn:PSD} with these parameter values produces results similar to those of the EAGLE, Illustris-TNG, and FIRE-II simulations \citep{Iyer.etal.2020}. An example of a galaxy SFH and corresponding PSD in such a model is presented in Figure~\ref{fig:PSD}. This validates our approach to generating stochasticity and demonstrates that a simple analytical model can produce SFR similar to the SFR in state-of-the-art galaxy formation simulations.

For each individual galaxy, we generate a sequence of $\Delta_{\rm MS}$ Gaussian random numbers spaced by $\Delta t=10\,\rm Myr$, as described in the previous section. This sequence is interpolated to a given $t$ during the integration of galaxy evolution. To account for the fact that the mean of a log-normal distribution is biased high compared to the unperturbed value, we adjust the mean of the $\Delta_{\rm{MS}}$ by subtracting the variance of the sequence divided by 2. 

Stochasticity is then introduced by perturbing the SFR outputs with $\Delta_{\rm{MS}}$:
\begin{equation}
\rm{SFR_{stoch}} = 10^{\Delta_{\rm{MS}}}\cdot\rm{SFR}
\end{equation}\label{eqn:stoch_SFR} 

We assume that the PSD does not change with redshift, since observed values of $\Delta_{\text{MS}}$ depend only mildly on redshift. We tested that varying $\Delta_{\text{MS}}$ by an amount significantly larger than the observed redshift dependence does not change our results.

Given that galactic outflows in the model are assumed to be proportional to SFR, $\dot{M}_{\rm out}=\eta \dot{M}_\star$ stochasticity should also induce stochasticity of the outflow rate. Physically, however, outflows may not necessarily respond to each local burst of star formation in a given star-forming region, but should arise from a collective effect of such local bursts. 
We thus assume that the outflow rate is proportional to the stochastic SFR rate averaged over 100 Myr time scale:
\begin{equation}
\dot M_{\rm{g,out}} = \eta_w \dot M_{\star,\rm{ave}},
\end{equation}
where $\eta_w$ is the mass-loading factor, and $\dot M_{\star,\rm{ave}}$ is the running average of the instantaneous SFR over the past 100 Myr.

%----------FIG: CMD
\begin{figure*}
	\includegraphics[width=\textwidth]{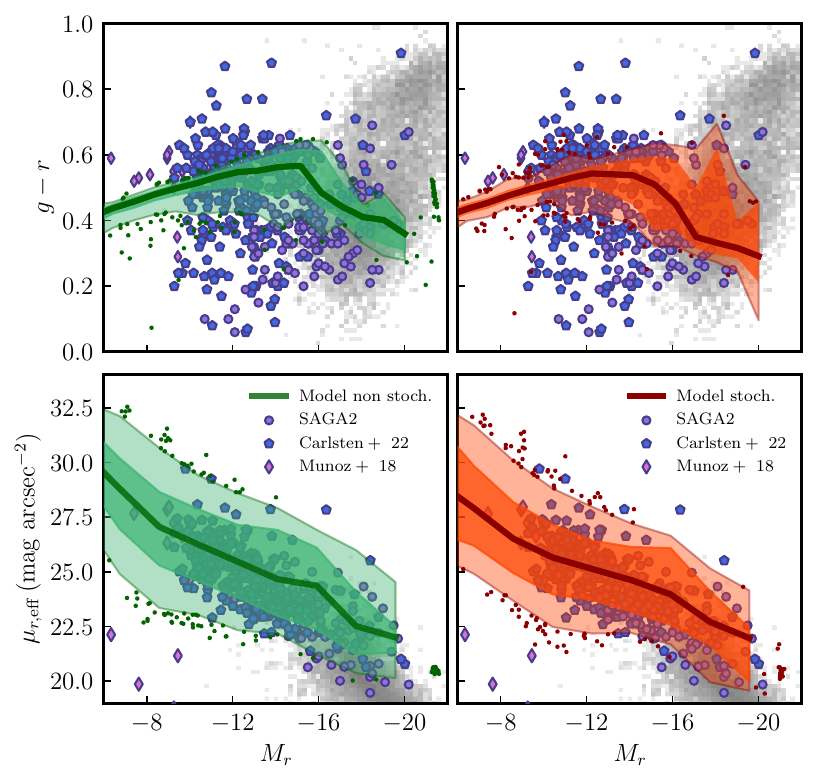}  
	\caption{\emph{Upper}: Colour--magnitude diagram of g-r colour vs $M_r$ relation for model galaxies in the non-stochastic (upper left) and stochastic runs (upper right) compared to observations such as the MegaCam survey \citet{Munoz.etal.2018}, the SAGA survey \citep{Geha.etal.2017, Mao.etal.2021}, and the ELVES survey \citep{Carlsten.etal.2020, Carlsten.etal.2022}. Similar to Figure~\ref{fig:SFR_mass}, we plot the median SFR for each mass bin (thick lines) and the 1-sigma and 2-sigma width with the 2-sigma outliers (green dots for the left two columns and red dots for the right two columns). We also show the distribution of dwarf galaxies in the GAMA survey as a 2D histogram in grey. The non-stochastic model exhibits a tightly clustered and almost linear colour-magnitude relation across magnitudes, which contrasts with observations that display more scatter in the $-20 < M_r < -8$ magnitude range. The addition of stochastic star formation rate (SFR) increases the scatter in CMD, particularly for brighter magnitudes $M_r < -12$, but improves the comparability with observations for galaxies brighter than $M_r = -12$, while still falling short of reproducing the comparable scatter in faint dwarf galaxies. \emph{Lower:} the $r$-band effective surface brightness as a function of $M_r$ for both non-stochastic and stochastic runs, with negligible differences in scatter between the two cases, and both able to reproduce the mean trend and scatter of the relation similar to observations. Further discussion on the findings is presented in Section~\ref{sec:discussion}.} \label{fig:CMD}
\end{figure*}

%----------FIG: CMD stoch Z
\begin{figure*}
	\includegraphics[width=\textwidth]{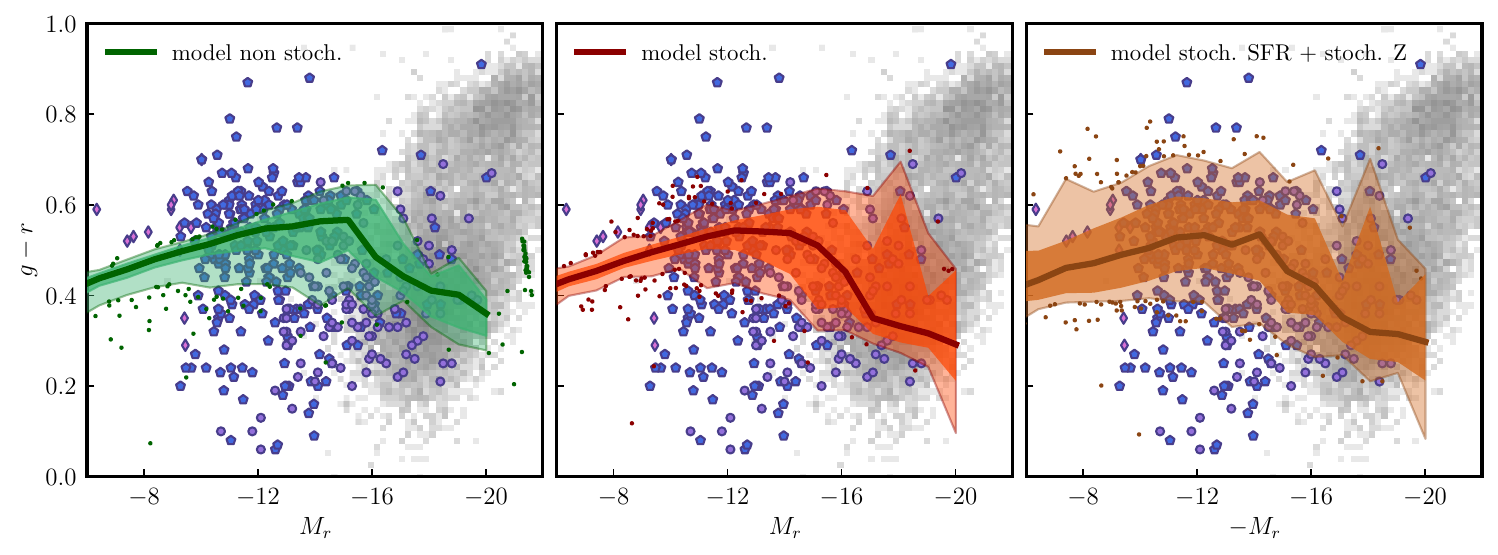}  
	\caption{Colour--magnitude diagram of g-r colour vs $M_r$ relation for model galaxies in the non-stochastic (left), stochastic SFR (middle), and stochastic SFR + stochastic metallicity runs (right) compared to the same observations as in Figure~\ref{fig:CMD}. Although perturbing the SFR does not increase the scatter in colour, especially for faint dwarf galaxies with $M_r > -12$, perturbing the metallicity significantly increases the scatter in colour across all magnitudes. This aligns with observations that show some quenched galaxies having blue colours and UV emission \citep{Greene.etal.2023}, where stochasticity in metallicity likely plays a role in their blue colour instead of star formation.} \label{fig:CMD_stoch_Z}
\end{figure*}

%----------FIG: mass vs metallicity 
\begin{figure*}
	\includegraphics[width=\textwidth]{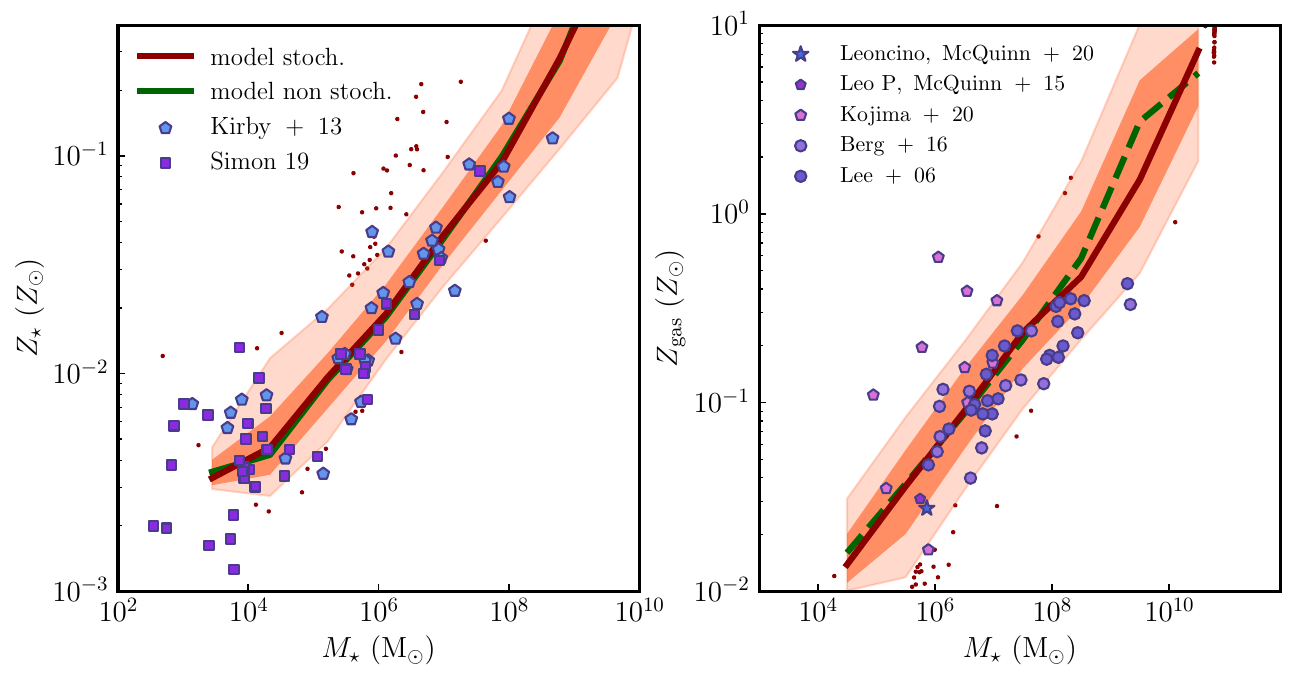}  
	\caption{The gas metallicity--mass (right) and stellar metallicity--mass (left) relations for the non-stochastic and stochastic runs with observations from various sources such as \citet{Lee.etal.2006, Berg.etal.2016, Kojima.etal.2020, McQuinn.etal.2015, McQuinn.etal.2020, Kirby.etal.2013, Simon.2019}. Similar to Figure~\ref{fig:SFR_mass}, we plot the median SFR for each mass bin and the 1-sigma width. In the non-stochastic case, the model accurately predicts the mean trend of the metallicity-mass relation, but the scatter is tighter than observed. The stochastic run shows slightly increased scatter, but it maintains the metallicity--mass relation without any breaks. The introduction of stochasticity in the SF model is seen to be effective in preserving the metallicity-mass relation.} \label{fig:metallicity_mass}
\end{figure*}

%----------SECTION 3: Effect of stochasticity on galaxy scaling relations
\section{Effect of stochasticity on galaxy scaling relations}
\label{sec:effect}
In this section, we explore effects of SFR stochasticity implemented described in the previous section on the  $M_\star-{\rm SFR}$, $M_{\rm g}-\rm SFR$ relations (Section~\ref{sec:effect_SFR}), and on the colour-magnitude relation (Section~\ref{sec:effect_color}).

%----------SUBSECTION 3.1: Effect of stochasticity on galaxy SFR
\subsection{Effect of stochasticity on galaxy SFR}\label{sec:effect_SFR}

Figure~\ref{fig:SFR_mass} shows a comparison of the $M_\star-\rm SFR$ and $M_{\rm g}-\rm SFR$ relation for model galaxies and observed galaxies from the ALFALFA survey \citet{Durbala.etal.2020} and a number of other samples \citep[][]{Jimmy.etal.2015,Teich.etal.2016,McGaugh.etal.2017,James.etal.2017,McQuinn.etal.2015a,McQuinn.etal.2019,McQuinn.etal.2021,Lin.etal.2022}. Compared to results shown in Figure 11 of \citet{Kravtsov.Manwadkar.2022} the widths of $M_\star-\rm SFR$ and $M_{\rm g}-\rm SFR$ relations are considerably larger for the model with added SFR stochasticity. However, the increase is not sufficient to account for extreme starbursting dwarfs reported by \citet{Lin.etal.2022} or even for galaxies with the highest SFR in other samples (e.g., SHIELD and STARBIRDS galaxies). 
This indicates that additional physical processes might be at play to boost the SFR to such a high level. One possibility is that SFR in the smallest dwarf galaxies in these samples is dominated by the formation and destruction of one or a couple of massive star clusters, that temporarily boost SFR by a large factor \citep[see, e.g.,][]{Zick.etal.2018}. 

We explore this possibility by adding an additional stochasticity to our model on the time scale of 10 Myr corresponding to a typical lifetime of a star-forming region that gives rise to individual clusters. This corresponds to adding the delta function to the PSD at this time scale. The amplitude of the delta function corresponds to the perturbation of the actual SFR of
\begin{equation}
    \rm{SFR_{\rm{pert.}}} = \rm{SFR_{\rm{stoch.}}} \cdot r \cdot 20 
\label{eqn:sfr_boost}
\end{equation}
where $r$ is a random variable drawn from the exponential distribution $3e^{-3x}$. 

The result is shown by the blue line and blue shaded region in Figure~~\ref{fig:SFR_mass}. Such additional stochasticity allows the model to match most observed galaxies, although not the most extreme starbursts. Such starbursts are very rare in observations, however, and only a dozen is found in the entire Dark Energy Survey.

Figure~\ref{fig:gas_stellar_mass} shows a comparison between non-stochastic and stochastic runs for the $M_g-M_\star$ relation. Introducing stochasticity does not change the mean trend and scatter of this relation. 

In Section~\ref{sec:dis_sfr_indicators}, we delve into the sensitivity of the stochasticity of SFR to the choice of SFR indicators, as demonstrated by previous studies \citep{Lee.etal.2009, daSilva.etal.2014, Emami.etal.2019}.

%----------SUBSECTION 3.2: Effect of stochasticity on galaxy color
\subsection{Effect of stochasticity on galaxy colours}\label{sec:effect_color}

Figure~\ref{fig:CMD} illustrates that the scatter in the $g-r$ colours for bright galaxies ($M_r < -12$)  modeled with additional SFR stochasticity agrees well with observation, whereas the scatter for fainter galaxies ($M_r > -12$) has a scatter much smaller than observed.  This discrepancy may be due to the fact that dwarf galaxies in the model remain quenched with after reionization and thus stochasticity does not affect their magnitudes.  Observed dwarf galaxies with masses $M_\star \gtrsim 10^5\, M_\odot$ often do show indications of recent star formation or star formation at intermediate epochs \citep[e.g.,][]{McQuinn.etal.2023,Jones.etal.2023}, and such rejuvenated dwarf galaxies are also present at least in some models of galaxy formation \citep[e.g.,][]{Rey.etal.2020}. 

To test the effect of reionization on the quenching of dwarf galaxies in the model, we lowered the critical mass. $M_c (z)$,  at which baryonic fraction is suppressed by a factor of 2 relative to the universal value due to post-reionization UV heating of the intergalactic medium (IGM). Thus,  lowering $M_c (z)$ results in more low-mass haloes that are able to accrete gas and prolong star formation after the epoch of reionization. However, lowering $M_c (z)$ by a factor of four does not produce significantly more blue low-mass dwarf galaxies. Changing the form of $M_c (z)$ to a step function with a filtering halo mass as low as $10^8 M_\odot$ also does not change the colour distribution for faint dwarf galaxies. Thus, a declining SFH and lack of scatter in colour for faint dwarf galaxies is not due to accretion suppression resulting from the post-reionization UV heating of the IGM.
Rather, it must be due to a reason internal to the galaxies themselves. 

One possibility is that the optical light of observed dwarf galaxies includes emission lines arising from ionized HII regions, while these are not included in calculations of colours of the model galaxies. Indeed,  \citet{Martin-Manjon.etal.2008} demonstrated that the presence of emission lines from the interstellar medium (ISM) can significantly impact the colours of starbursting dwarf galaxies (see their Figures 7 and 10). To assess this effect, we employed the Flexible Stellar Population Synthesis (\texttt{FSPS}) code \citep{Conroy.etal.2009,Conroy.etal.2010} in conjunction with its Python bindings, \textsc{PyFSPS} \footnote{\href{https://github.com/dfm/python-fsps}{\tt https://github.com/dfm/python-fsps}}, to generate single stellar population instances. We compared the colours of these instances with and without incorporating the contribution of nebular emission lines and found only minor differences. Therefore, the difference in the scatter of colours is unlikely to be due to the omission of nebular emission lines in the colour calculation. 

Recently, \citet{Greene.etal.2023} showed that many early-type dwarf galaxies ($M_\star \lesssim 10^7 M_\odot$) exhibit excess UV emissions and blue colours. These blue colours are not due to ongoing star formation, but due to significant variations of metallicity between galaxies of a given $M_\star$. Lower metallicity galaxies have bluer colours and scatter in metallicity thus should result in the scatter of colours. To test this, we perturbed each model galaxy's metallicity $Z$ by a log-normal Gaussian random number:
\begin{equation}
    Z_{\rm{stoch}} = Z \cdot 10^{\Delta_{\rm{Z}}},
\end{equation}
where $\Delta_{\rm{Z}}$ is a Gaussian random number centered on 0 with standard deviation of $\sigma=0.5$. 

The effect of such additional metallicity scatter on colours is shown in the right panel of Figure~\ref{fig:CMD_stoch_Z}. Comparison of the colour--magnitude relations in the three panels of this figure shows that additional scatter in metallicity indeed leads to a significant increase in scatter of colour, especially for dwarf galaxies with $M_r > -12$. Thus, the tail of blue colours of quenched low-mass dwarf galaxies in observations could be due to metallicity scatter. At the same time, as we show below, inclusion of such large metallicity scatter overestimates the scatter in the metallicity--stellar mass relation.

%----------SUBSECTION 3.3: Effect of stochasticity on metallicity-mass relation 
\subsection{Effect of SFR stochasticity on the stellar mass-metallicity relation}\label{sec:effect_feh}

Given the potentially significant effects of metallicity scatter on dwarf galaxy colours we investigate the impact of stochastic SFH on the stellar mass--metallicity relation of dwarf galaxies, which is widely utilized as a diagnostic tool for modeling galaxy feedback, chemical evolution, inflows, and outflows \citep[e.g.][]{Garnett.2002, Peeples.Shankar.2011, De.Lucia.etal.2020, Finlator.Dave.2008}. In the case of dwarf galaxies, observations have been conducted to measure stellar metallicity using absorption lines in stellar emission \citep[see, for instance, the review by][]{Simon.2019} and to determine gas metallicity using emission lines from HII regions in galaxies with gas and ongoing star formation \citep{Lee.etal.2006, Berg.etal.2012, Berg.etal.2016}.

Figure~\ref{fig:metallicity_mass} illustrates the relation between galaxy stellar mass and gas and stellar phase metallicities, $Z_{\rm g}$ and $Z_\star$, of model and observed dwarf galaxies. The scatter observed in the $Z_\star-M_\star$ and $Z_g-M_g$ relations is comparable or slightly higher in the stochastic run when compared to the non-stochastic run. 

The width of the 1-sigma band indicates that the overall scatter in the gas metallicity--gas mass relationship is higher than the scatter in the stellar metallicity--stellar mass relationship. This can be attributed to the fact that $Z_\star$ is the average metallicity in the cumulative stellar mass and increases monotonically, whereas $Z_{\rm g}$ is influenced by star formation and metal enrichment outflows and inflows, which govern the instantaneous gas mass \citep{Finlator.Dave.2008, Lilly.etal.2013, Torrey.etal.2019, van.Loon.2021}. It is worth noting that the model $Z_{\rm g}$ and $Z_\star$ values do not consider the observational uncertainties that contribute to the scatter of the relations of observed galaxies.

\begin{figure}
	\includegraphics[width=\columnwidth]{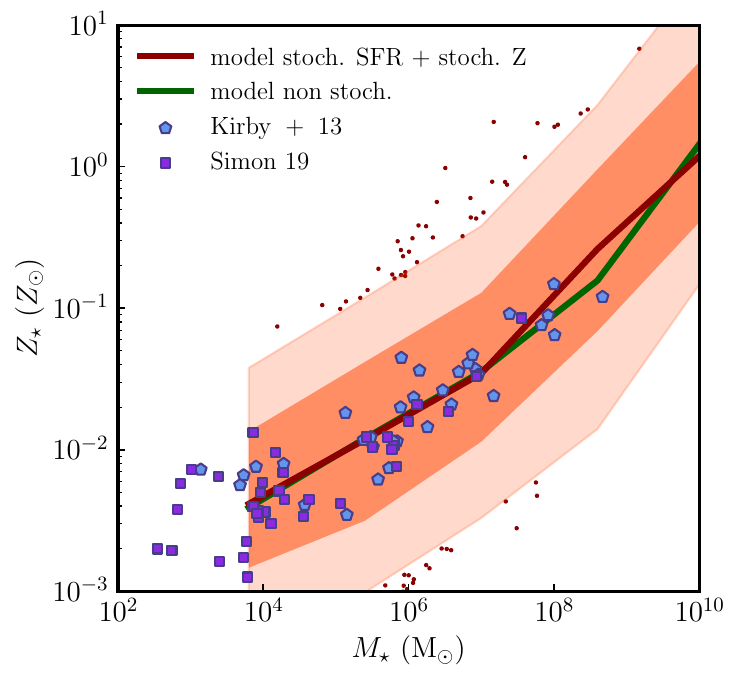}  
	\caption{The stellar metallicity--mass relation for the non-stochastic and stochastic SFR + stochastic metallicity runs compared with the same observations as in Figure~\ref{fig:metallicity_mass}. Similar to Figure~\ref{fig:metallicity_mass}, we plot the median $Z_\star$ for each mass bin and the 1-sigma and 2-sigma width. The stochastic SFR + stochastic metallicity run shows an overestimated scatter compared to observations.} \label{fig:metallicity_mass_stoch_Z}
\end{figure}

Figure~\ref{fig:metallicity_mass_stoch_Z}, on the other hand, shows that additional stochasticity in the metallicity explored in the previous subsection (eq.~\ref{fig:metallicity_mass_stoch_Z}) results in a metallicity scatter that is significantly larger than observed. 

%----------SECTION 4: Discussion
\section{Discussion}\label{sec:discussion}

Results presented above indicate that star formation rates in dwarf galaxies are very stochastic. Furthermore, we find that adding SFR stochasticity to the SFHs of model galaxies results in increased $g-r$ colour scatter in bright dwarf galaxies $(M_r < -12)$ bringing their colour distribution in a better agreement with observations. In contrast, colours of faint dwarf galaxies $(M_r > -12)$ are largely unaffected by the addition of the SFR stochasticity. 

In this section, we discuss interpretation of these results, compare our results with those of other theoretical models and observational studies, examine how observed SFR stochasticity depends on the use of different SFR indicators, and investigate the link between short-term and long-term fluctuations and galaxy evolution. Lastly, we highlight the limitations of interpretation of our results and outline potential avenues for future research.

%----------SUBSECTION 4.1: Comparison with other models
\subsection{Comparison with other models}\label{sec:dis_comp_models}

It is interesting to compare the level of SFR stochasticity required to match observed scatter in the $M_\star-\rm SFR$ relation in the dwarf galaxy regime with stochasticity in various galaxy formation models and simulations presented by  \citet{Iyer.etal.2020}. These authors computed the PSD of galaxies with $10^9 M_\odot < M_\star < 10^{11.5} M_\odot$ in cosmological  simulations (Illustris \citep{Genel.etal.2014,Vogelsberger.etal.2014}, Illustris TNG \citep{Weinberger.etal.2017,Pillepich.etal.2018}, Mufasa \citep{Dave.etal.2016}, Simba \citep{Dave.etal.2019}, EAGLE \citep{Crain.etal.2015,Schaller.etal.2015,Schaye.etal.2015}), zoom-in simulations (FIRE-2 \citep{Hopkins.etal.2018}, g14 \citep{Brooks.Zolotov.2014,Brooks.Christensen.2016,Christensen.etal.2016,Christensen.etal.2018,Brooks.etal.2017}, Marvel/Justice League) \citep{Bellovary.etal.2019}, semi-analytical models \citep{Somerville.etal.2008,Porter.etal.2014,Somerville.etal.2015,Brennan.etal.2017} and semi-empirical UniverseMachine model \citep{Behroozi.etal.2013,Behroozi.etal.2019}. 

\citet{Iyer.etal.2020} found that different models exhibit a wide range of SFR stochasticity. The largest stochasticity levels are reached in the simulations with efficient, explosive feedback that do not rely on subgrid ISM models, such as EAGLE and FIRE-2 simulations. Simulations that do use a subgrid ISM model exhibit considerably less stochasticity on time scales $\lesssim 100$ Myr. Semi-analytic models that do not model local ISM structure and do not add stochasticity explicitly, such as the Santa Cruz SAM and Universe Machine, exhibit lower SFR stochasticity still. This is because the SFR stochasticity arising due to variations in the mass assembly histories and internal global parameters of galaxies are sub-dominant to the stochasticity on $\lesssim 100$ Myr time scales induced by the rise and fall of individual star-forming regions in galaxies \citep{Tacchella.Forbes.Caplar.2020,Shin.etal.2023}. 

The inherent level of stochasticity in the \texttt{GRUMPY} model is comparable to that of simulations with subrid physics and semi-analytic models. On the other hand, the addition of explicit SFR stochasticity at the level required to match observed scatter in the $M_\star$-SFR relation brings the SFR stochasticity to the level similar to that exhibited by low-mass galaxies in the FIRE-2 and EAGLE simulations. 

%----------SUBSECTION 4.2: Color scatter comparison
\subsection{Scatter of the $g-r$ colours compared to observations and simulations}\label{sec:dis_color_scatter}

The main challenge that we identified is reproducing a large observed scatter of $g-r$ colours of faint dwarf galaxies ($M_r\gtrsim -15$). 
\citet{Chaves_Montero.2021} examined effects of star formation rate burstiness on the colours of massive galaxies ($\log_{10} M_\star>9.25$) in the Illustris TNG simulations and model galaxies in the UniverseMachine. They found that burstiness can affect optical colours of individual galaxies at the $\approx 0.2-0.3$ level, but does not significantly change the colour distribution of galaxy population as a whole. This is different from our finding in the dwarf galaxy regime, likely because the level of burstiness implied by the observed scatter in the SFR-$M_\star$ regime is significantly larger than the level considered in that study. 

Comparison of observations with the Auriga simulation suite of MW-sized galaxies \citep[][]{Grand.etal.2017} revealed that the colour-magnitude distribution for $M_\star\lesssim 10^6\, M_\odot$ in simulations is considerably narrower than in observations, and there are notably fewer faint blue dwarf galaxies in simulations than in observations \citep{Pan.etal.2023}.  These authors attributed this discrepancy to the absence of low-temperature/molecular gas cooling in the stellar feedback and ISM model used in the Auriga simulations. In the case of faint dwarf galaxies, the molecular gas self-shielding effect is likely underestimated when the UV radiation perturbs the gas due to the limitations of the model. Therefore, most smallest dwarf galaxies are quenched and become red, with almost no star-forming blue galaxies forming following reionization. 

This is similar to what we find in the \texttt{GRUMPY} model for smaller galaxies. The reasons may also be similar as star formation in this model is based on the model of molecular hydrogen of \citet{Gnedin.Draine.2014}, which was not tested at the metallicities of the ultra-faint galaxies. At the same time, star formation in low-metallicity galaxies may not be well traced by molecular hydrogen to begin with \citep[][Polzin et al. in prep.]{Nobels.etal.2023}. Thus, the star formation rate may be underestimated in the faintest dwarf galaxies due to the adopted star formation model in \texttt{GRUMPY}.

Field galaxies of $M_\star \gtrsim 10^6\, M_\odot$ in both Auriga and APOSTLE simulations continue forming stars to low redshifts, although satellite galaxies do quench after their accretion onto MW-sized system \citep{Digby.etal.2019}. Likewise, \citet{Garrison_Kimmel.etal.2019} also find that galaxies with $M_\star\lesssim 10^6\, M_\star$ are quenched by UV heating after reionization in the FIRE-2 zoom-in simulations of dwarf galaxies. However, they also find that galaxies of larger mass generally continue forming stars until $z=0$.

In the GRUMPY model galaxies of larger mass also continue to form stars and their colours are affected by SFR stochasticity as was demonstrated above.

%----------SUBSECTION 4.3: Sensitivity of stochasticity on different SFR indicators
\subsection{Contribution of uncertainties in observational SFR indicators to the apparent SFR stochasticity}\label{sec:dis_sfr_indicators}

Although there are physical drivers of bursty star formation in dwarf galaxies, part of the apparent scatter of the SFR in these galaxies could be due to uncertainty in the observational SFR indicators. Different observed bands contain distinct information about star formation, with shorter wavelengths probing most recent star formation, while longer wavelengths probing star formation over longer time scales \citep[e.g.,][]{Boquien.etal.2014}. Additionally, spectral features such as $H \alpha$ and UV-based SFR indicators, $H \alpha$ and Ca-H,K absorption line equivalent widths, the D$_n$(4000) spectral break, and broadband galaxy colours are sensitive to the model parameters in the PSD formalism \citep{Iyer.etal.2022}. Therefore, comprehending the variability of SFRs across different indicators can provide information about stochasticity on different time scales and inform comparisons with model results.

The most commonly used SFR indicators are $H \alpha$ and UV+IR fluxes \citep{Kennicutt.1998, Kennicutt.Evans.2012, Boquien.etal.2014, Davies.etal.2016}. $H \alpha$ photons originate from gas ionized by young stars ($<$20 Myr), providing a measure of the current SFR $(\sim 4-10)$ Myr in galaxies. In dwarf galaxies, however, 
H$_\alpha$ can significantly underpredict the star formation rate due to paucity of massive stars \citep[e.g.,][]{Lee.etal.2009}. UV emission directly comes from star-forming regions, while some is absorbed and reprocessed by dust, re-emerging in the IR. Thus, the sum of UV and total IR luminosities provides the total SFR over the timescale of $\approx 30-100$ Myr.

The distribution of SFRs measured using two indicators can be used to investigate the change in SFRs over time and provide a statistical view of the behavior of the galaxy population. However, such analysis is challenging due to several uncertainties and assumptions, including the stochastic IMF and dust properties, the monotonicity of SFRs over different timescales, and the inherent difficulty in measuring SFRs accurately \citep{Johnson.etal.2013, Shivaei.etal.2018}.

At low SFRs, inferring an SFR from a star formation indicator (SFI) in an individual galaxy is problematic. When calculating the sum of the mass and ages of all the stars that comprise a population using stellar population synthesis (SPS) models to map observed luminosity to underlying physical properties, it is common to assume that the IMF and SFH are well populated. However, if the IMF is not fully sampled, the luminosity's highly non-linear dependence on stellar mass can have significant consequences for the population's luminosity. Thus, there is no longer a deterministic relation between the total mass and age of the population to the total luminosity and colour of its integrated light. This makes the inverse problem of determining the mass or age of a simple stellar population from its photometric properties ill-posed.  When the IMF and SFH are sparsely sampled in the low SFR regime, there is no unique mapping between SFRs and SFIs, and thus no unique way to infer an SFR from an SFI in an individual \citep{daSilva.etal.2014}.

For SFRs $\lesssim 10^{-4} M_\odot$/yr, the scatter in the $H\alpha$ indicator, which is used for most dwarfs at these masses, is expected to be very high due to insufficient sampling of IMF and SFH. In addition, \citet{Lee.etal.2009} demonstrated that for lower luminosity dwarf galaxies (roughly less active than the Small Magellanic Cloud), H$\alpha$ tends to increasingly underpredict the total SFR relative to the FUV. By SFR $\sim 0.003 M_\odot$ yr$^{-1}$, the average H$\alpha$-to-FUV flux ratio is lower than expected by a factor of two, and at the lowest SFRs probed, the ratio exhibits an order of magnitude discrepancy for the handful of galaxies that remain in the sample due to the IMF deficiency in dwarf and low surface brightness galaxies. The upper panels in Figure 4 of \citet{daSilva.etal.2014} illustrate these results. According to \citet{Cignoni.etal.2019}, the level of stochasticity on a 10 million-year time scale is quite significant, whereas it becomes much less significant on a 100 million-year time scale.

%----------SUBECTION 4.4: CAVEATS
\subsection{Caveats}\label{sec:caveats}
The interpretation of our results should be approached with caution due to the constraints of our simple regulator-type model.

First, we make the assumption of a smooth, radial-averaged exponential profile for both the gas surface density $\Sigma_g$ and the molecular gas surface density $\Sigma_{\rm{HI}}$. However, in reality, the interstellar medium (ISM) can be highly structured \citep{Kravtsov.2003,Semenov.etal.2017}, and a radial average may not accurately capture the overdensities and underdensities within the ISM. This is particularly relevant for low-mass galaxies, where star formation occurs in a small number of bright star-forming regions, making it difficult to effectively average out fluctuations. In blue compact dwarf galaxies, some giant molecular clouds are massive enough to produce significant torques on surrounding disk material, leading to kinematic decoupling and stirring up the inhomogeneity of the ISM. Thus, a closer examination of the detailed structure of the ISM is necessary in future studies.

Second, we applied the same level of stochasticity to all dwarf galaxies with stellar masses between $10^5 M_\odot$ and $10^9 M_\odot$. However, previous studies have demonstrated that the intrinsic scatter of the main sequence (MS) for star-forming galaxies with stellar masses between $10^8 M_\odot$ and $10^{11} M_\odot$ does not remain constant, but exhibits a "U-shape" trend \citep{Davies.etal.2022, Huang.etal.2023}. Specifically, for low-mass galaxies with $10^8 M_\odot < M_\star < 10^{10} M_\odot$, stochastic starbursts and stellar feedback events are the main drivers of intrinsic MS scatter. For intermediate-mass galaxies around $10^{10} M_\odot$, the effect of star formation and feedback becomes less pronounced, resulting in a decrease in intrinsic MS scatter. In contrast, high-mass galaxies ($M_\star > 10^{10.3} M_\odot$) exhibit an increased intrinsic MS scatter due to AGN feedback. For even lower-mass dwarf galaxies with $M_\star < 10^8 M_\odot$, the star formation rate is subject to small-number statistics and becomes more bursty. Therefore, to accurately model these variations, the level of stochasticity incorporated in the model should be a function of stellar mass, with higher stochasticity for low-mass galaxies and decreasing stochasticity as we move towards intermediate masses around $10^{10} M_\odot$. However, as this project is a proof of concept, we defer a more detailed study of the dependence of stochasticity on galaxy stellar mass to future work, focusing instead on demonstrating the non-negligible impact of incorporating stochasticity on several galaxy scaling relations.

%----------SECTION 5: Conclusions
\section{Conclusions}\label{sec:conclusion}
We present the impact of bursty star formation on several galaxy scaling relations in dwarf galaxies using the \texttt{GRUMPY} model. This simple regulator-type model reproduces the average trend of the SFR--mass, stellar mass--gas mass, stellar mass--metallicity relation, and CMD. However, the scatter of these relations in the original model is smaller than observed. To address this, we utilize a formula for the power spectral density (PSD) to generate a series of random numbers with a log-normal distribution to perturb the original SFR. We adjust the parameters in the PSD formalism to control the amplitude and variability of burstiness in the model. We experiment with the effects of long-term and short-term SFR variations in the model and compare them to observations. The main results can be summarized as follows:
\begin{enumerate}
    \item We find that the PSD of equation~\ref{eqn:PSD} with parameters $\alpha=2$, $\tau_{\rm{break}}=2$Gyr, and $\sigma = 0.6$ produces both short-term and long-term variations, and the amplitude of burstiness  comparable to zoom-in galaxy formation simulations, such as FIRE-2. 
    \item Bursty star formation increases the scatter in the SFR-$M_\star$ relation compared to the non-bursty case. We find, however, that
    rare extreme starburst galaxies, such as those in \citet{Lin.etal.2022} imply even larger levels of stochasticity. 
    \item Adding the SFR stochasticity  increases the scatter in the CMD (Figure~\ref{fig:CMD}) in brighter dwarf galaxies $(M_V < -12)$. For fainter dwarf galaxies $(M_V > -12)$, the scatter is comparable to that of the non-bursty model. We confirmed that this result is not due to low-mass dwarf galaxies being quenched by reionization. 
    \item We show that metallicity scatter can increase the scatter of colours of faint dwarfs $(M_V > -12)$, however this leads to the overestimate of scatter in the mass-metallicity relation compared to observed Local Group dwarf satellite galaxies.

\end{enumerate}

%----------ACKNOWLEDGEMENT
\section*{Acknowledgements}
This research made use of data from the SAGA Survey (saga- survey.org), which was supported by NSF collaborative grants AST-1517148 and AST-1517422 and by Heising-Simons Foundation grant 2019-1402. 

The software packages NUMPY \citep{van.der.Walt.etal.2011}, SCIPY \citep{Jones.etal.2001}, Astropy \citep{astropy:2018}, MATPLOTLIB \citep{Hunter.2007}, and GITHUB were invaluable in conducting the analyses presented in this paper. Additionally, this research relied heavily on the Astrophysics Data Service (\href{https://ui.adsabs.harvard.edu/classic-form/}{\tt{ADS}}) and the preprint repository \href{https://arxiv.org/}{\tt{arXiv}}.

We thank Anirudh Chiti and Nick Gnedin for constructive feedback on an earlier version of this paper.

%----------DATA AVAILABILITY
\section*{Data Availability}
The data underlying this article will be shared on reasonable request to the authors.

%----------REFERENCES

\bibliographystyle{mnras}
\bibliography{main}

\bsp
\label{lastpage}
\end{document}